\documentclass[twocolumn]{autart}    

\usepackage{graphicx}          

\begin{document}

\begin{frontmatter}

\title{Relationship between Fragility, Diffusive Directions and Energy
       Barriers in a Supercooled Liquid} 


\author[UAM]{Manuel I. Marqu\'es}\ead{manuel.marques@uam.es},    
\author[BU]{H. Eugene Stanley}\ead{hes@bu.edu},               

\address[UAM]{Departamento de F\'isica de Materiales C-IV,
Universidad Aut\'onoma de Madrid, 28049 Madrid, Spain}            
\address[BU]{Center for Polymer Studies and Department of Physics,
Boston University, Boston, MA 02215, USA}                         

\begin{keyword}                           
Supercooled liquids; Fragility; Glasses; Energy Lansdcape .               
\end{keyword}                             

\begin{abstract}                          
An analysis of diffusion in a supercooled liquid based solely in the
density of diffusive directions and the value of energy barriers shows
how the potential energy landscape (PEL) approach is capable of
explaining the $\alpha$ and $\beta$ relaxations and the fragility of a
glassy system. We find that the $\beta$ relaxation is directly related
to the search for diffusive directions. Our analysis shows how in strong
liquids diffusion is mainly energy activated, and how in fragile liquids
the diffusion is governed by the density of diffusive directions. We
describe the fragile-to-strong crossover as a change in the topography
of the PEL sampled by the system at a certain crossover temperature
$T_\times$.
\end{abstract}

\end{frontmatter}

\section {Introduction}

The study of slow dynamics in disorder systems and, in particular, the
study of the glass transition in supercooled liquids is a topic of
considerable interest in condensed matter physics. In general, liquids
are divided in two different classes depending on the properties of
their glass transitions, strong and fragile \cite{Angell}. Strong
liquids experience a gentle increase in the relaxation upon cooling,
often according to the Arrhenius law, close to the glass transition
temperature $T_{g}$.  On the other hand, fragile liquids, experiment a
sharp rise of the viscosity, increasing by several orders of magnitude
in a very narrow interval of temperature close to $T_{g}$. The glass
transition temperature is not related to any dynamical transition and is
experimentally defined as the one where the value of the viscosity is
$10^{13}$ P. For strong liquids, nothing special happens close to
$T_{g}$ and the glass transition shows a conventional behavior where no
transition may be defined. However, fragile systems seem to present a
divergence behavior close to $T_{g}$ indicating that some kind of new
dynamical mechanism may be responsible for the onset of the glassy
phase. Theoretical attempts conducted to study this possible new
transition have been mainly focused in two different approaches: the
mode coupling theory (MCT) \cite{Geszti}, and the potential energy
landscape (PEL) \cite{Goldstein}.

MCT studies structural arrest in supercooled liquids as a purely dynamic
singularity happening as a result of a feedback between shear-stress,
diffusion and viscosity. The idealized MCT predicts structural arrest to
take place at a temperature $T_{\mbox{\scriptsize MC}}>T_{g}$
\cite{Bengtzelius,Gotze}. To restore ergodicity below
$T_{\mbox{\scriptsize MC}}$ additional hopping or activated mechanism
have been introduced into the theory \cite{GotzeII}, avoiding the
kinetic singularity. MCT accurately describes important aspects of
relaxation dynamics in liquids above their melting temperatures, in
particular, the behavior of the intermediate scattering function and the
$\alpha$ and $\beta$ relaxations. The $\beta$ relaxation plateau is
related to the time expended by the particle to break the cage formed by
neighboring particles. The accuracy of the MCT predictions above $T_{g}$
has been verified experimentally and using computer simulations
\cite{GotzeIII,Kob,KobII}.

The potential energy landscape (PEL) approach has been studied using the
concept of an inherent structure (IS) \cite{Stillinger} (i.e., the
configurations at the local minimum of the system's potential
energy). Several numerical and theoretical studies have provided
evidence of a relation between the dynamics of the supercooled liquid
and the PEL. It was found that correlation functions display stretching
in time in the same temperature range in which the systems explores
local minima of the PEL with deeper and deeper energy
\cite{Sastry,Buchner}, a thermodynamic description of the supercooled
liquid was performed in terms of the IS configurational entropy
\cite{Sciortino}, fragility was related to properties of the PEL
\cite{SastryII} and the diffusion process was analyzed in terms of the
visited inherent structures \cite{Angelanib}.

There was no connection between both theories (MCT and PEL). Basically
there has not been a clear definition of $T_{\mbox{\scriptsize MC}}$
from the potential energy landscape point of view and a precise
landscape-based definition of hopping and activated dynamics has been
lacking \cite{Debenedetti}. This situation has recently changed due to
the instantaneous normal mode approach to the PEL \cite{Keyes}. This
approach relates the diffusive processes to the number of accessible
paths in the multidimensional energy landscape
\cite{Donati,LaNave,Bembenek}. In particular, a key point is the
temperature dependence of the fraction of negative eigenvalues of the
Hessian calculated at the saddle points of the PEL
\cite{Angelani,Broderix}. It was found that this fraction approaches
zero at $T_{\mbox{\scriptsize MC}}$, but is appreciable at larger
temperatures.  Since the negative eigenvalues correspond to diffusive
directions on the PEL, the following scenario was proposed for the
dynamics of supercooled liquids close to the glass transition: For
$T>T_{\mbox{\scriptsize MC}}$ the system lies close to saddles in the
configuration space and the relevant dynamic process is a diffusion
among the saddle points along paths at almost constant potential energy,
where there is no need to overcome any energetic barrier (``border
dynamics'').  That implies that the factors impeding free diffusion of
the particles are related to finding these paths (entropic factors),
rather than due to energetic factors. However, below
$T_{\mbox{\scriptsize MC}}$, there are few paths available and diffusion
is dominated by hopping processes allowing the system to evolve from
minimum to minimum (minimum-to-minimum dynamics).  This implies that a
sharp slowing down on the dynamics should take place close to
$T_{\mbox{\scriptsize MC}}$ if the energetic barriers to be crossed by
the system are high enough \cite{Grigera}.

\section{Analytic approach to the PEL}

To check if this proposed mechanism is correct we are going to study the
problem analytically, relating the properties of the PEL---the density
of diffusive directions and value of the energy barriers---directly to
the fragile and strong characteristics of supercooled liquids, and to the
$\alpha$ and $\beta$ relaxations predicted by MCT.

We consider particles following a Brownian motion in three-dimensional
space.  In the absence of any other kind of interaction the mean square
displacement is given by
\begin {equation}
\langle r^{2}(t)\rangle=6T[t-(1-e^{-t})].
\label{eq1}
\end{equation}
Here we set the Boltzmann constant $k=1$, and take the particle masses
$m=1$. For very short times ($t\ll 1$), the particles behave as free
particles $\langle r^{2}(t)\rangle=3Tt^{2}$, and for longer periods of
time ($t\gg 1$) the particles behave as diffusive particles in a random
walk $\langle r^{2}(t)\rangle=6Tt$.

For particles in supercooled liquids such as Lennard-Jones systems,
silica or water, the evolution is much more complicated due to
non-trivial interactions among particles. These non-trivial interactions
produce a very complicated energy landscape, making analytical results
very difficult to obtain. However, a lot of information about these
energy landscapes has been obtained using powerful numerical techniques
such as Molecular Dynamics or Monte Carlo simulations. In particular, it
has been proposed that the density of diffusive directions with
temperature, $k(T)$, follows a power law \cite{Angelani}
\begin {equation}
k(T)=A(T-T_{\mbox{\scriptsize MC}})^\gamma.
\label{eq2}
\end{equation}
The typical value $\Delta E$ of the energy barriers in hopping processes
close to the mode coupling critical temperature, has been determined for
different models \cite{Grigera,AngelaniII}. In all cases it has been
found that a good approximation is given by
\begin {equation}
\Delta E \approx 10T_{\mbox{\scriptsize MC}}.
\label{eq3}
\end{equation}
For long enough time (we consider a time $t$ to be long enough if
$t>1$), the diffusion of a particle is no longer free, making the total
system to evolve over the multidimensional PEL. From a thermodynamic
point of view the diffusion of the particle takes place if the direction
chosen on the PEL turns to be a diffusive one, or if the barrier is low
enough to be overcome by means of an activated process. Considering
Eqs.~(\ref{eq2}) and (\ref{eq3}), the probability of diffusion of the
particle at a temperature T from a time $t>1$ to a time $t+1$ is given
by
\begin {equation}
P_{\mbox{\scriptsize diff}}=A(T-T_{\mbox{\scriptsize
    MC}})^{\gamma}+[1-A(T-T_{\mbox{\scriptsize MC}})^\gamma]e^{-\Delta
  E/T}.
\label{eq4}
\end{equation}
Considering $P_{\mbox{\scriptsize diff}}$, it is possible to relate the
total value of the mean square displacement $\langle R^{2}(t)\rangle$ (at
time $t>1$) with the value of the random walk diffusion $\langle
r^{2}(t)\rangle$ by
\begin {equation}
\langle R^{2}(t)\rangle=\sum_{i=1}^{t} {t-1 \choose i-1}\langle
r^{2}(i)\rangle P_{\mbox{\scriptsize diff}}^{i-1}(1-P_{\mbox{\scriptsize
diff}})^{t-i}.
\label{eq5}
\end{equation}
If $t<1$ the diffusion is the one given by a random walk, $\langle
R^{2}(t)\rangle=\langle r^{2}(t)\rangle$.

\section {Results for $\langle R^{2}(t)\rangle$}

Next, we study $\langle R^{2}(t)\rangle$ from Eq.~(\ref{eq5}) for four
different cases. Our standard case is going to be a binary mixture
Lennard-Jones (BMLJ) system with density $\rho=1.2$, as the one studied
in ref.\cite{Angelani} for which $T_{\mbox{\scriptsize MC}}=0.435$,
$\gamma=0.94$ and $A\approx0.05$.  We will study four cases.

\begin{itemize}

\item Case (a): We consider the simplest system where only Brownian
motion is present and there is no effect of the PEL. We take $k(T)=1$,
implying that any direction chosen by the system is going to be a
diffusive one.

\item Case (b): The second system considered has a larger density of
diffusive directions that the one studied in \cite{Angelani}. To do so
we take $A\approx 0.5$. In this particular case $k(T)=1$ for
$T>2.5$. 

\item Case (c): We consider the standard case from ref.\cite{Angelani}
where $k(T)\not=1$ for every $T$ in this study.

\item Case (d): We consider a case where there are no diffusive
directions $A=0 \Rightarrow k(T)=0$.

\end{itemize}

\begin{figure}
\includegraphics[width=7cm,height=7cm,angle=0]{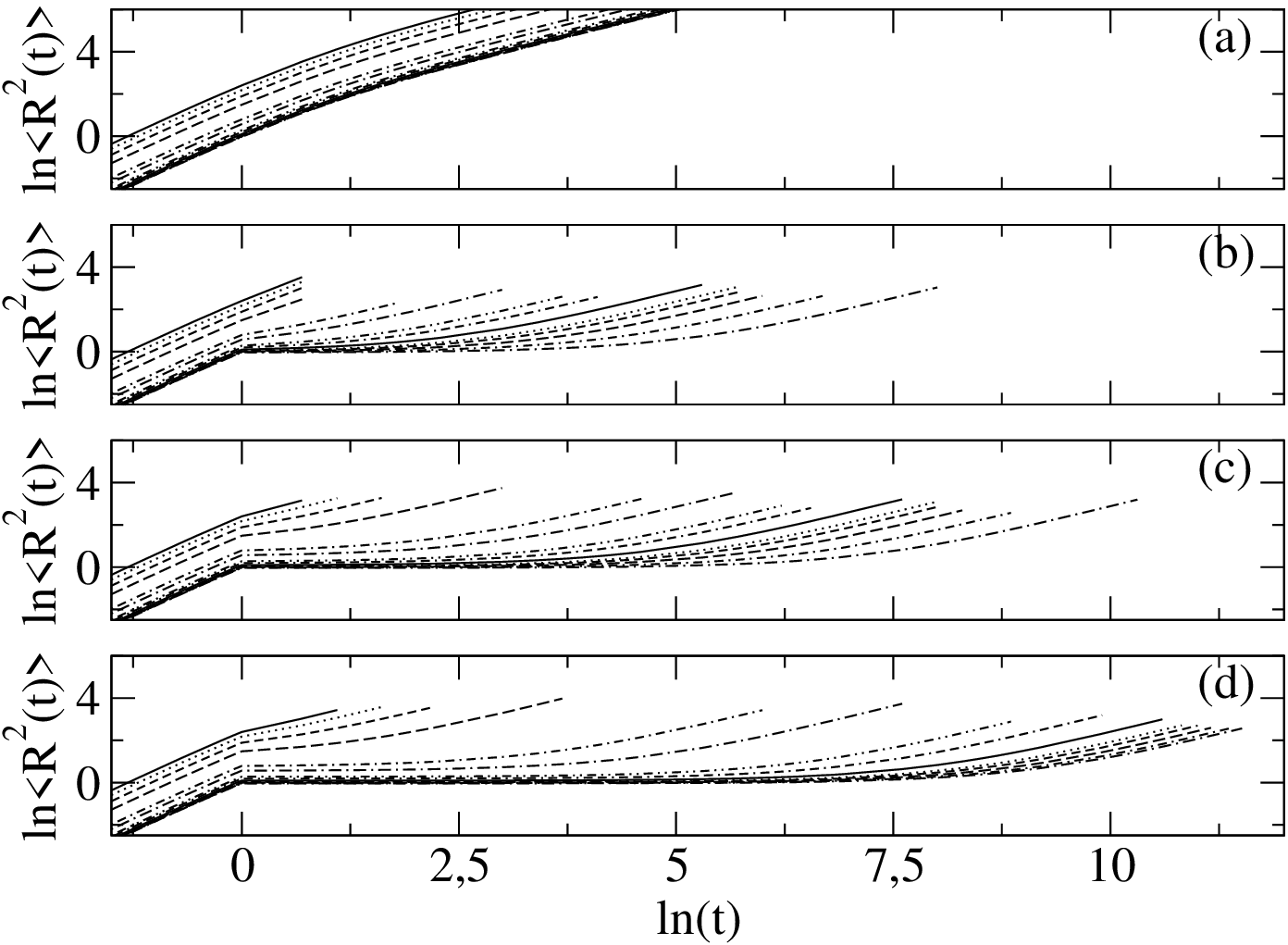}
\caption{$\ln\langle R^{2}(t)\rangle$ vs. $\ln(t)$ obtained from
Eq.~(\ref{eq5}) for temperatures (from top to bottom) $T=5$, 4, 3, 2, 1,
0.8, 0.6, 0.55, 0.5, 0.475, 0.466, 0.454, 0.444, and 0.439. The values
of the parameters are $\gamma=0.94$, $T_{\mbox{\scriptsize MC}}=0.435$,
$\Delta E=10T_{\mbox{\scriptsize MC}}$ and (a)$k(T)=1$, (b)$A \approx
0.5$, (c)$A\approx0.05$ and (d)$A=0$.  }
\label{fig1}
\end{figure}

Results are presented in Fig.~\ref{fig1} For temperatures ranging from
$T=5$ to $T=0.4386$.  A clear relation between the $\beta$ relaxation
time and the density of diffusive directions in the system, $k(T)$, is
found in Fig.~\ref{fig1}. In Case (a) all directions are diffusive and
particles do not expend any $\beta$ relaxation time searching for a
diffusive direction to scape, that is the reason why there is no plateau
in Fig.~\ref{fig1}(a). On the contrary, in Case (d), there are no
diffusive directions and the particle expends a long time in the $\beta$
relaxation plateau. The only possible diffusion mechanism to scape from
this plateau is by an activated process to overcome the $\Delta E$
barrier. So the $\beta$ relaxation time may be interpreted in terms of
the PEL as a ``search'' for diffusive directions.  This mechanism should
be equivalent to the one described in the MCT where the particles
``search'' for directions to scape from the cage formed by surrounding
particles.

\section{Results for $D(T)$}

Once $\langle R^{2}(t)\rangle$ is known it is possible to determine the
diffusion coefficient $D(T)$ as a function of temperature, considering
that a straight line, with unit slope, fitted to the long time behavior
of data in Fig.~\ref{fig1}, intersects a vertical line at $\ln(t)=0$ at a
height $\ln(6D)$. To study what is the mechanism in the PEL leading to
the behavior of a supercooled liquid as strong or fragile, we study
three different cases.

\begin{itemize}

\item Case (a): We study a system with no diffusive directions, where
the only possibility for the system to diffuse is an activated process
to overcome the energetic barrier $\Delta E \simeq
10T_{\mbox{\scriptsize MC}}$. We consider again a value $A=0 \Rightarrow
k=0$.

\item Case (b): We consider the opposite case, where no activation
process is available (since the energy barriers are extremely high) and
the only diffusion mechanism is the search for diffusive directions. In
order to study a system like that we take $\Delta E \simeq
10^{5}T_{\mbox{\scriptsize MC}}$ and $A\approx 0.05$.

\item Case (c): Finally we consider the Binary Lennard Jones system
considered in \cite{Angelani} where both diffusive mechanisms are
available.

\end{itemize}

\begin{figure}
\includegraphics[width=7cm,height=7cm,angle=0]{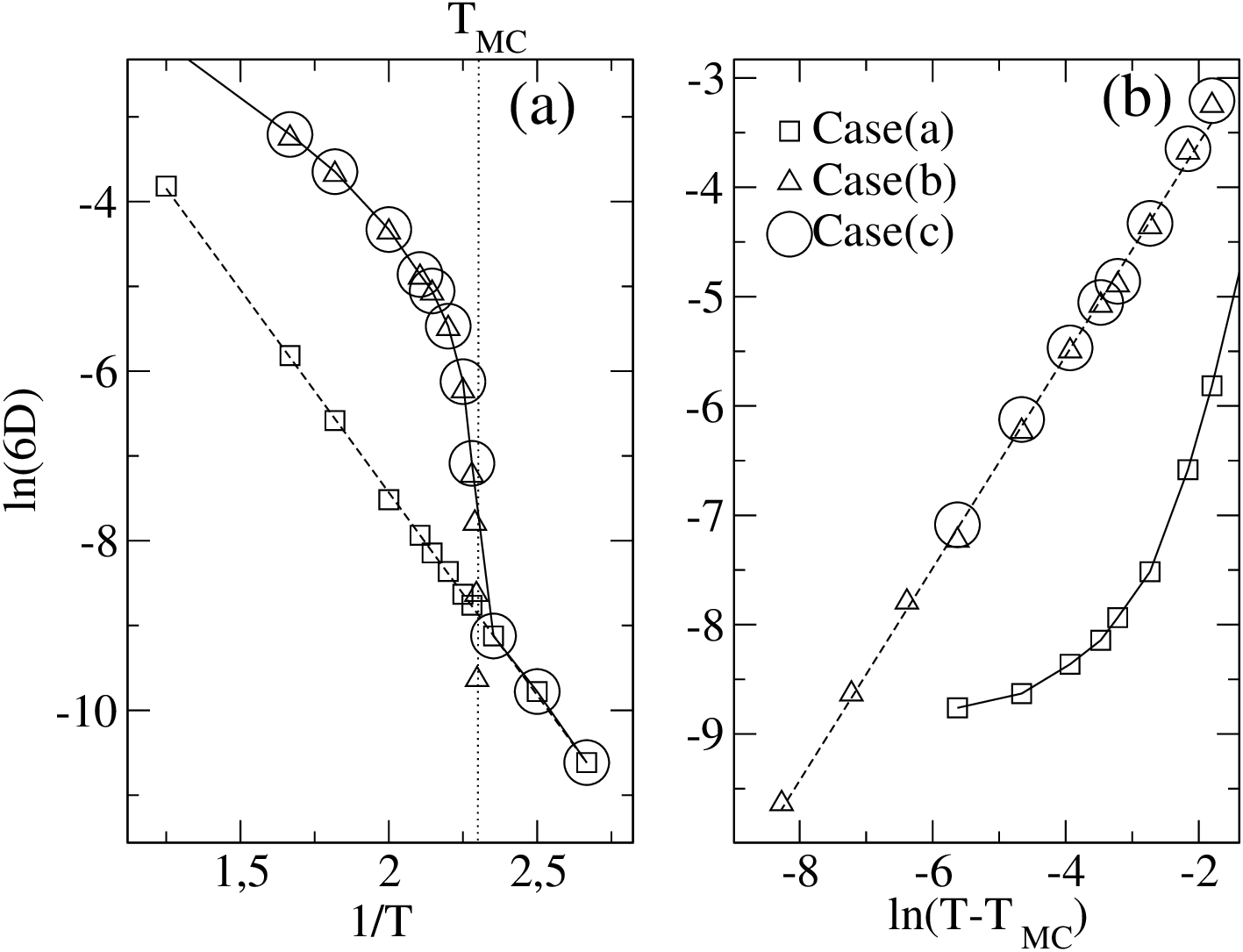}
\caption{Figure 2(a) shows $\ln(6D)$ vs. $1/T$. Dashed line is a liner
fit to the data for Case (a), dotted line marks the value
$T=T_{\mbox{\scriptsize MC}}$. Figure 2(b) shows
$\ln(6D)$ vs. $\ln(T-T_{\mbox{\scriptsize MC}})$. Dashed line is a liner
fit to the data for Case (b). The values of the parameters in the
different cases are Case (a) $\Delta E=10T_{\mbox{\scriptsize MC}}$,
$A=0$, Case (b) $\Delta E=10^{5}T_{\mbox{\scriptsize MC}}$,
$A\approx0.05$ and Case (c) $\Delta E=10T_{\mbox{\scriptsize MC}}$,
$A\approx0.05$.}
\label{fig2}
\end{figure}

Results are presented in Fig.~\ref{fig2}. We plot $\ln(6D)$ vs. $1/T$ in
Fig.~\ref{fig2}(a) and $\ln(6D)$ vs. $\ln(T-T_{\mbox{\scriptsize MC}})$
in Fig.~\ref{fig2}(b). Note how Case (a) clearly shows a linear (Arrhenius
behavior), typical of a strong liquid. Nothing special happens when
$T=T_{\mbox{\scriptsize MC}}$, since the mechanisms of diffusion are
always the same (activated processes).  On the contrary Case (b) presents
the typical behavior of a fragile liquid predicted by MCT. In this case,
the transition to a glassy state at $T=T_{\mbox{\scriptsize MC}}$, is a
singular one. Since the only mechanism of diffusion are the diffusive
directions and those are equal to zero at $T=T_{\mbox{\scriptsize MC}}$
the system gets trapped in a glass state by dynamical arrest. Note how
the barrier must be very high to get a really slowing down in the
dynamics, making any kind of hopping impossible. This result agrees with
predictions made in \cite{Grigera}. If the energy barriers are not high
enough we are in Case (c) and we have a transition from strong to fragile
behavior at $T=T_{\mbox{\scriptsize MC}}$ as the one predicted by
\cite{Cavagna} and observed numerically in \cite{LaNaveII}. Note how
Case (c) is almost identical to Case (b) when the temperatures are far
from $T_{\mbox{\scriptsize MC}}$ clearly indicating that the mechanisms
of diffusion are governed mostly by ``border dynamics,'' that is, by
evolution of the system through the diffusive direction of the
PEL. However for $T<T_{\mbox{\scriptsize MC}}$, Case (c) behaves like
Case (a), which means that the mechanisms of diffusion are now governed
by activated processes to overcome the $\Delta E$ energy
barriers. Case (c) is a clear example of crossover from ``border dynamics''
to ``minimum-to-minimum dynamics.''

Results in Fig.~\ref{fig2} are easy to understand considering that,
since $P_{\mbox{\scriptsize diff}}$ is the probability for a particle to
diffuse, a $P_{\mbox{\scriptsize diff}}\not=0$ turns $t$ into a lower
value, $P_{\mbox{\scriptsize diff}}t$. That means that instead of
$\langle R^{2}(t)\rangle=6Tt$ for $t>>1$, we have $\langle
R^{2}(t)\rangle=6TP_{\mbox{\scriptsize diff}}t$. For the Case (a) (strong
glass), since $A=0$, we have $P_{\mbox{\scriptsize diff}}=e^{-\Delta
E/T}$, making $\langle R^{2}(t)\rangle=6Te^{-\Delta E/T}t$, and getting a
value of the diffusion coefficient $\ln(D)=\ln(T)-\Delta E/T$.  For the
Case (b) (fragile glass) it is possible to consider in good
approximation that $P_{\mbox{\scriptsize diff}}=A(T-T_{\mbox{\scriptsize
MC}})^{\gamma}$, obtaining
$\ln(D)=\ln(T)+\ln(A)+\gamma\ln(T-T_{\mbox{\scriptsize MC}})$, which is
the behavior expected by MCT. The constant value for $\Delta E$ given in
Eq.~(\ref{eq3}) is valid only near $T_{\mbox{\scriptsize MC}}$. If we
want to obtain results from our model in a larger range of temperatures,
a non-constant value of $\Delta E$, as the one reported in
\cite{Grigera}, should be taken into account.

\section{Crossover from Fragile to Strong}

Numerical simulations and theoretical calculations have recently shown that the density of diffusive
directions is not exactly zero at $T_{\mbox{\scriptsize MC}}$
\cite{Fabricius,Shell}. It has been argued that this density behaves as an
Arrhenius exponential decay which is only zero at $T=0$
\cite{Doliwa}. Actually, a close inspection of the data reported in
Ref.~\cite{LaNaveII} shows clearly that $k(T)$ is almost null at
$T_{\mbox{\scriptsize MC}}$, but not exactly zero.

If we change the power law behavior in Eq.~(\ref{eq2}) by an exponential
decay given by
\begin {equation}
k(T)=A^{*}e^{-\Delta E^{*}/T},
\label{eq6}
\end{equation}
with $A^{*}$ constant and $\Delta E^{*}$ an energy scale, we obtain
\begin {equation}
P_{\mbox{\scriptsize diff}}=A^{*}e^{-\Delta E^{*}/T}+e^{-\Delta
E/T}-A^{*}e^{-(\Delta E^{*}+\Delta E)/T}.
\label{eq7}
\end{equation}
Equation~(\ref{eq7}) implies an Arrhenius behavior for $D(T)$, making
impossible to find a crossover from fragile to strong.  It means that
there must be some crossover temperature $T_\times$ where $k(T)$ changes
from a power-law to an Arrhenius-law, marking a change in the PEL
topography and the beginning of a crossover from fragile to strong.
Experimental results \cite{Rossler} have shown that $T_\times$ normally
has a value very close to $T_{\mbox{\scriptsize MC}}$.

To study the effect of this possible change on topography we are going
to modify the density of diffusive directions of Case (C), considering a
new Case (D) with
\begin{eqnarray}
\nonumber k(T)=&f_{0}e^{\Delta E[(1/T_{\mbox{\scriptsize MC}})-(1/T)]}
& \qquad T<T_\times, \\
k(T)=&A(T-T_{\mbox{\scriptsize MC}})^{\gamma}
& \qquad T>T_\times, 
\end{eqnarray}
where $f_{0}$ is the density of diffusive directions at
$T_{\mbox{\scriptsize MC}}$, which is now set to a value different from
zero ($f_{0}=0.001$). The crossover temperature, $T_\times$, is given by
the lower root of the equation
\begin{equation}
f_{0}e^{\Delta E[(1/T_{\mbox{\scriptsize
MC}})-(1/T_\times)]}=A(T_\times-T_{\mbox{\scriptsize MC}})^{\gamma}.
\label{eq9}
\end{equation}
A plot of $k(T)$ vs. $T$ is shown in Fig.~3a compared with the one
corresponding to Case (C), where $k(T_{\mbox{\scriptsize MC}})$ is
strictly equal to zero.

Results for $\ln(6D)$ vs. $1/T$ are given in Fig.~3b and compared to the
behavior of the pure fragile system. Now the crossover from fragile to
strong takes place at $T=T_\times$ (close to $T_{\mbox{\scriptsize
MC}}$) and it is not so abrupt as the one in Fig.~2a. The qualitative
behavior shown in Fig.~3b coincides with the result found for Silica by
means of Molecular Dynamics Simulations \cite{LaNaveII}. However, the
behavior reported in Fig.~2a is more similar to the one found for BMLJ
\cite{AngelaniII}, posibly indicating that for BMLJ $f_{0} \approx 0$
and $T_{\mbox{\scriptsize MC}} \approx T_\times$.

\begin{figure}
\includegraphics[width=7cm,height=7cm,angle=0]{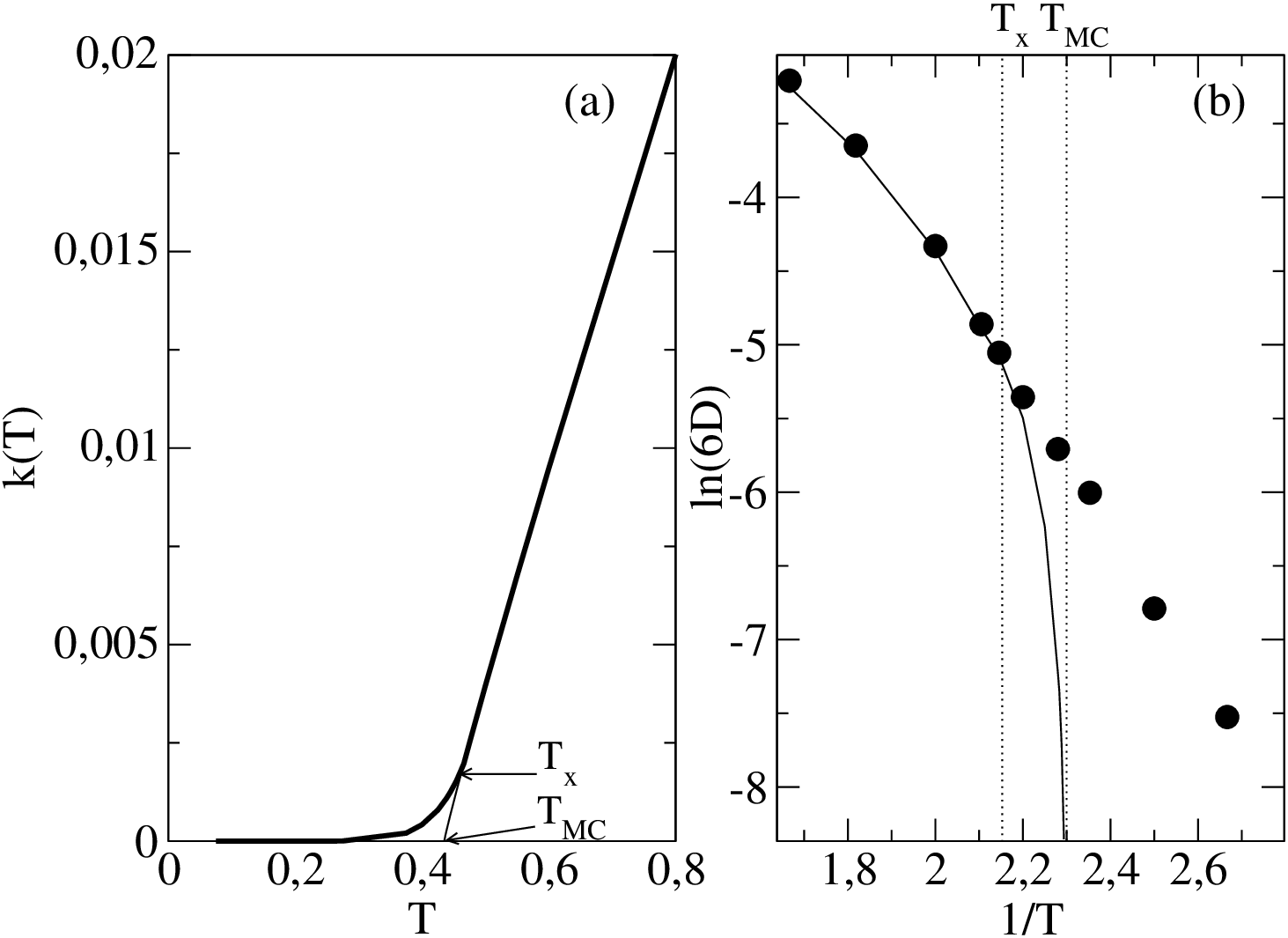}
\caption{Figure 3(a) Density of diffusive directions $k(T)$
vs. temperature $T$ for Case (C) (thin line) and Case (D) (thick
line). $T_{\mbox{\scriptsize MC}}$ marks the value $k(T)=0$ for Case (c)
and $T_\times$ marks the change on topography of the PEL for Case
(D). Figure 3(b) shows $\ln(6D)$ vs. $1/T$ for Case (D) (black
points). Continuous line represents the behavior for the pure fragile
glass former. $T_{\mbox{\scriptsize MC}}$ and $T_\times$ are marked with
dotted lines.}
\label{fig3}
\end{figure}

\section{Conclusions}

To conclude, we have analyzed the diffusion in a supercooled liquid
based solely on the density of diffusive directions and activated
processes and have shown how the PEL provides an explanation for the
$\beta$ and $\alpha$ relaxations and the fragility characteristics of a
glassy system. The $\beta$ relaxation is directly related to the
attempts of the system to move through the diffusive directions. The PEL
shows that a ``strong'' liquid is one in which the main mechanisms of
diffusion are ``activated dynamics'' and that a ``fragile'' liquid
exhibits dynamics typical of supercooled liquids with diffusive
directions but very high barriers where hopping is almost impossible.
In this case PEL supports the same dynamical arrest behavior predicted
by MCT. The crossover from fragile to strong is found to be related to a
change on the topography of the PEL at a certain crossover temperature
$T_\times$, where the density of diffusive directions changes from
power-law to Arrhenius.

\begin{ack}                               
We would like to thank F. Sciortino and N. Giovambattista for useful
discussions, and the Spanish Ministry of Education and the NSF Chemistry
Program for support.  
\end{ack}

\bibliographystyle{plain}        
\bibliography{autosam}           

\end{document}